\documentclass[12pt]{article}

\usepackage{amsmath}
\usepackage{graphicx}
\usepackage{multirow}
\usepackage{epsfig}
\usepackage{rotating}

\newcommand{\ba}{\begin{array}}
\newcommand{\ea}{\end{array}}

\begin{document}

\begin{flushright}
DCPT-09-164\\
DFTT 65/2009\\
IPPP-09-82\\
SHEP-09-25\\
\today
\end{flushright}
\vspace*{0.8truecm}

\begin{center}
{\LARGE \bf Explicit CP violation in the MSSM Higgs sector}

\vspace*{0.8truecm}
{\large S. Hesselbach$^a$\footnote{stefan.hesselbach@durham.ac.uk},
S. Moretti$^{b,c}$\footnote{stefano@soton.ac.uk},
S. Munir$^d$\footnote{smunir@fisica.unam.mx},
P. Poulose$^e$\footnote{poulose@iitg.ernet.in}}

\vspace*{0.5truecm}
{\it
$^a$ IPPP, University of Durham, Durham, DH1 3LE, UK\\
$^b$ School of Physics \& Astronomy, University of Southampton,\\
 Highfield, Southampton SO17 1BJ, UK\\
$^c$ Dipartimento di Fisica Teorica, Universit\`a di Torino, Via Pietro Giuria 1,\\
 10125 Torino, Italy\\
$^d$ Instituto de F\'{i}sica, Departmento de F\'{i}sica Te\'orica, Universidad Nacional\\ Aut\'onoma de M\'exico, Apartado Postal 20-364, 01000, M\'exico, D.F.\\
$^e$ Physics Department, IIT Guwahati, Assam 781039, INDIA}
\end{center}

\begin{abstract}
\footnotesize
\noindent We analysed the sensitivity of the process $gg \rightarrow H_1 \rightarrow \gamma \gamma$ to the explicitly CP-violating phases $\phi_\mu$ and $\phi_{A_f}$ in the Minimal Supersymmetric Standard Model (MSSM) at the Large Hadron Collider (LHC), where $H_1$ is the lightest Supersymmetric Higgs boson. We conclude that depending on these phases, the overall production and decay rates of $H_1$ can vary up to orders of magnitude compared to the CP-conserving case.\\

\textbf{PACS:} 4.80.Cp Non-standard-model Higgs bosons, 12.60.Jv Supersymmetric models

\end{abstract}

\section{Introduction}

In the MSSM,
the Higgs potential conserves Charge \& Parity
(CP) at tree level \cite{Higgs-hunter}. Beyond the 
latter, CP violation can manifest itself through complex Yukawa couplings of the Higgs bosons to 
(s)fermions. There are several new parameters 
in the SUSY theory that are absent in the SM, which could well be 
complex and thus possess CP-violating phases. However, the CP-violating phases associated with the sfermions of the first and, to a 
lesser extent, second generations are severely constrained by bounds on the 
Electric Dipole Moments (EDMs) of the electron, neutron and muon \cite{EDM1}--\cite{EDM4}. 

By building on the results of 
Refs.~\cite{dedes1,dedes2} (for the production) 
and \cite{Hesselbach:2007gf}--\cite{Moretti:2007th} (for the decay), we recently examined the LHC phenomenology of the 
$gg\to H_1\rightarrow \gamma\gamma$ process (where
$H_1$ labels the lightest neutral Higgs state of the CP-violating MSSM), which involves the (leading) direct effects of CP violation through couplings of the $H_1$ to sparticles in the loops as well
as the (subleading) indirect effects through scalar-pseudoscalar mixing yielding the CP-mixed state $H_1$. Here we summarize the results of \cite{SMunir} focussing especially on the effects of a light stop in the production of a cp-mixed $H_1$ by gluon fusion and its decay into two photons. 

\section{CP violation in the di-photon search channel}

\begin{figure}[h!]
\centering \includegraphics[scale=0.75, clip=]{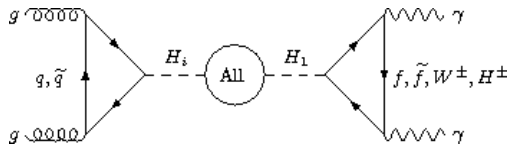}
\caption{
Leading order
Feynman diagram for $gg\rightarrow H_1\rightarrow \gamma\gamma$ including the 
effect of mixing in the propagator.
}
\label{fig:feyn}
\end{figure}

Explicit CP violation arises in the Higgs sector of the MSSM when various related 
couplings become complex. As a result, the physical Higgs bosons 
are no more CP eigenstates, but a mixture of them. CP-violating effects in the combined production and decay process enter through:
\begin{enumerate}
\item Complex $H_1$-$\tilde f$-$\tilde f^*$ couplings at production level.
\item Complex $H_1$-$\tilde f$-$\tilde f^*$ couplings at decay level.
\item Mixing in the propagator.
\end{enumerate}
The leading contribution to Higgs production in gluon fusion
is at the one loop level.  Similarly the leading contribution to di-photon decay
channel is also at one loop level, as shown in Fig. \ref{fig:feyn}.

The propagator is considered in the following way.
A CP-mixed Higgs particle, $H_i$, produced through gluon fusion, 
can be converted into another mass eigenstate, $H_j$, through loops of 
fermion or gauge boson (see Fig. \ref{fig:feyn}). The lightest of the three $H_j$ states is taken to be $H_1$ here. 
This $H_1$ decays then through the di-photon channel. The propagator matrix
is obtained from the self-energy of the Higgs particles computed at one-loop
level, where we used the expressions provided by \cite{Ellis:2004fs}
which include off-diagonal absorptive parts.
The matrix inversion required is done numerically using the Lapack \cite{lapack}
package. All the relevant couplings and masses are obtained from CPSuperH 
vesrsion 2 \cite{cpsuperh2}. Cross section amplitude 
of the full process shown in Fig. \ref{fig:feyn}, and the parton level cross 
section itself are then computed numerically. 

It was observed in \cite{Moretti:2007th} that the only significant contribution to the cross section is made by the phase of a light stop in the decay loops. Therefore, assuming a similar trend for the production mode also, we have considered a few sample parameter space
points and studied the effect of CP violation, in particular the significance 
of a light stop in the loops, ignoring the effects of other sfermions, in order to illustrate the typical effects of CP-violation in the MSSM. 

We fix the following MSSM parameters which do not play a role in CP
violation studies:\\

\(M_1=100~{\rm GeV},~~~M_2=M_3=1~TeV,~~~M_{Q_3}=M_{D_3}=M_{L_3}=M_{E_3}=M_{\rm SUSY}=
1~{\rm TeV}.\)\\

We consider the case of all the third generation tri-fermion couplings being 
unified into one single quantity, $A_f$.
All the soft masses are taken to be at some unification scale,
whose representative value adopted here is 1 TeV. When
considering the light stop case we take a comparatively light value
for $M_{U_3}\sim 250$ GeV, which corresponds to a stop mass of around 200 GeV,
 otherwise $M_{U_3}$ is set to 1 TeV. In the Higgs scalar-pseudoscalar coupling the
product of $\mu A_f$ and the sum of their phases is relevant rather than $\mu$ or $A_f$ separately. Hence, we have kept $\phi_{A_f}=0$ and
studied the effect of CP violation by varying $\phi_\mu$ alone.  
In our numerical analysis we have varied these parameters between 1 TeV and
2 TeV. $M_{H^+}$ is varied between 100 and 300 GeV. The mass of the
lightest Higgs particle is then in the range of 100--130 GeV. 
It was noticed that only large $\tan\beta$ case produce significant differences, so we take a 
representative value of $\tan\beta=20$ to see the effect of the
other parameters. 

\section{Results}

\begin{figure}[h!]
\centering 
\begin{tabular}{cc}
\hspace{-20pt}\includegraphics[scale=0.35]{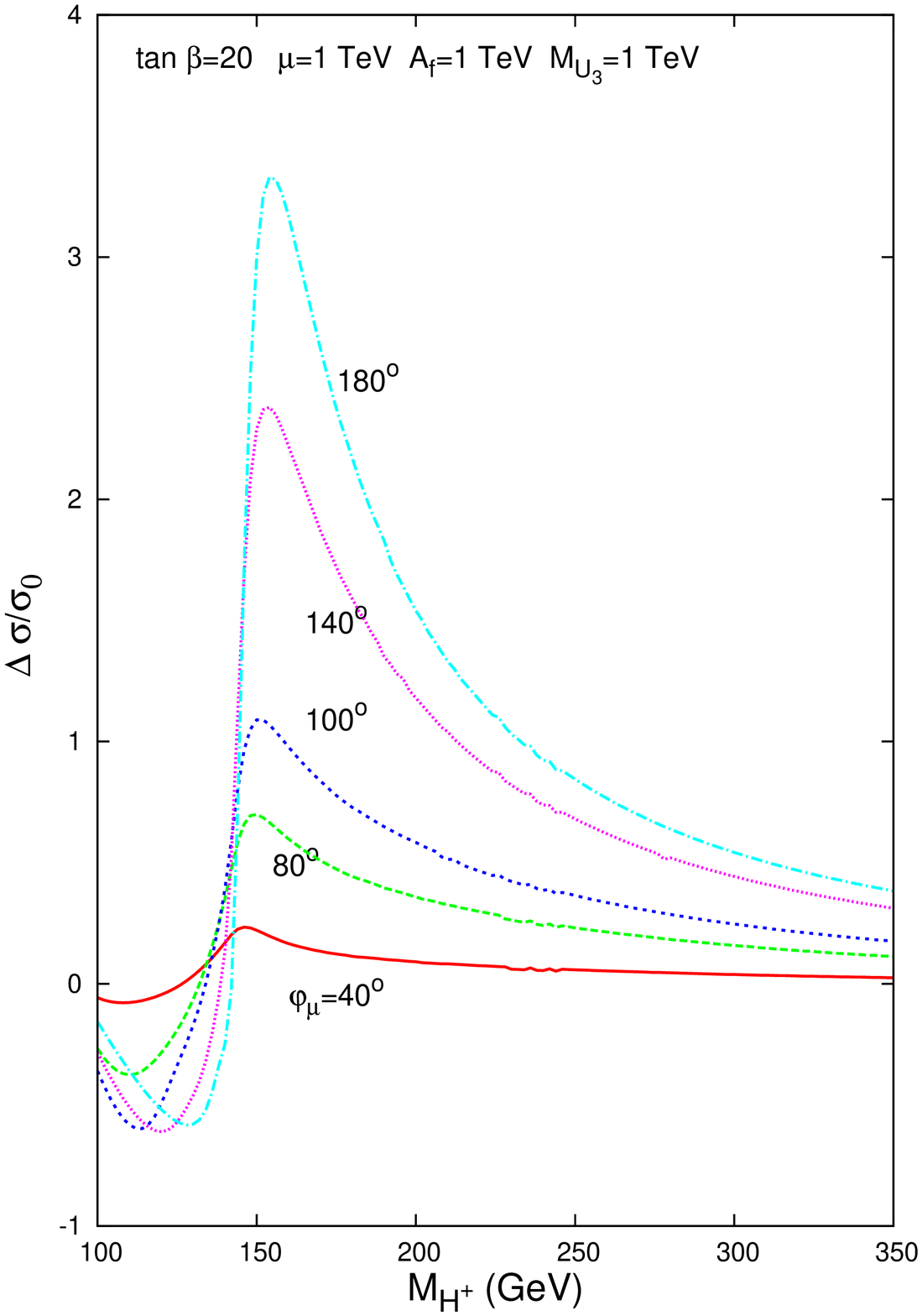} & \hspace{30pt} \includegraphics[scale=0.35]{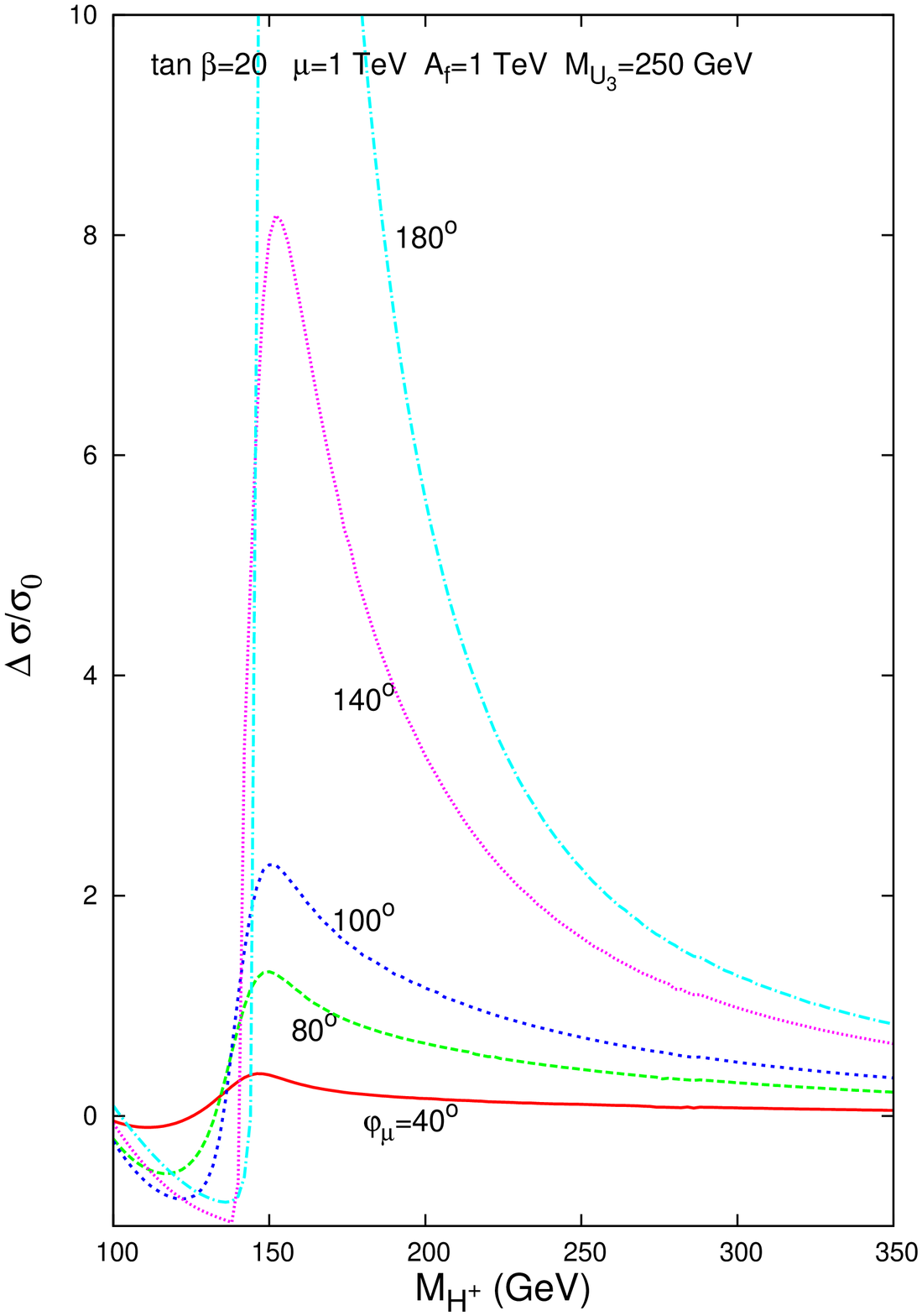} 
\end{tabular}
\caption{$\Delta\sigma /\sigma_0$ for the process $(pp\rightarrow H_1 \rightarrow \gamma\gamma)$
at the LHC plotted against the charged
Higgs mass for different $\phi_\mu$ values. The relevant MSSM variables are
set as follows: $\tan\beta=20$, $A_f=1$ TeV, $\mu=1$
TeV, with $M_{U_3}=1$ TeV (left plot) and $M_{U_3}=250$ GeV (right plot).} 
\label{fig:tb20A1mu1phimu}
\end{figure}

In order to show the dependence on phase, we plotted a quantity
$\Delta\sigma / \sigma_0$, with $\Delta \sigma$ being the difference
in the magnitude of cross section for $\phi_\mu$ set to a given value
and $\phi_\mu = 0$, and $\sigma_0$ being the cross section for $\phi_\mu = 0$, in Fig.~\ref{fig:tb20A1mu1phimu}, against $M_{H^+}$. We have considered $\mu=1$~TeV and $A_f=1$~TeV.  Clearly, there is appreciable variation of the cross section with $\phi_\mu$. Comparing the two cases of light and heavy stops, it is clear
that the Higgs-stop-stop coupling is significant, with the difference in cross sections being noticably large with a light stop.

\begin{figure}[h!]
\centering 
\begin{tabular}{cc}
\hspace{-20pt}\includegraphics[scale=0.35]{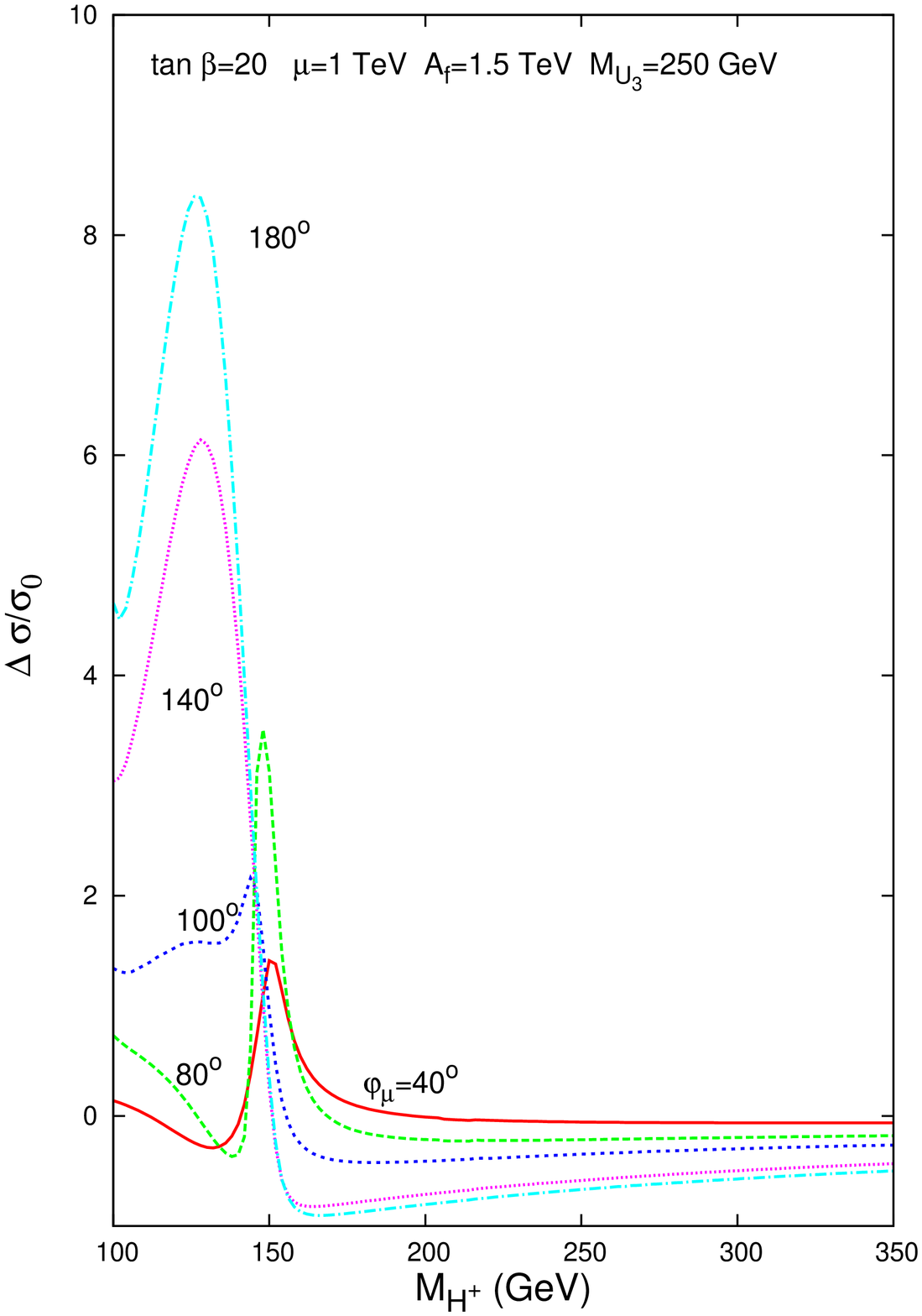} & \hspace{30pt} \includegraphics[scale=0.35]{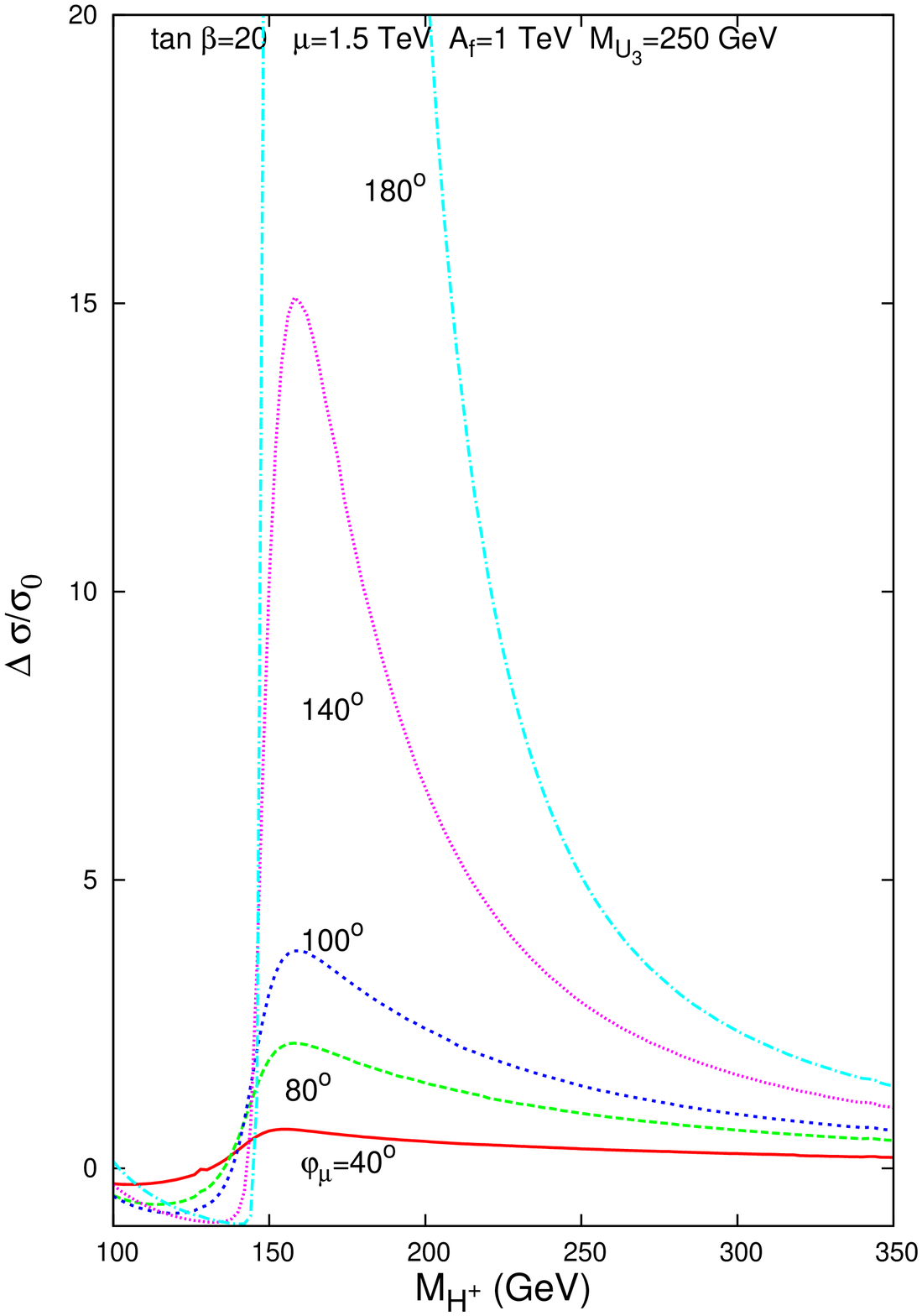} 
\end{tabular}
\caption{$\Delta\sigma /\sigma_0$ for the process $(pp\rightarrow H_1 \rightarrow \gamma\gamma)$
at the LHC plotted against the charged Higgs mass for different $\phi_\mu$ values, with $A_f=1.5$ TeV and $\mu=1$
TeV (left plot) and $A_f=1$ TeV and $\mu=1.5$
TeV (right plot), for $M_{U_3}=250$ GeV.} 
\label{fig:tb20A1mu1phimu3}
\end{figure}

Figure \ref{fig:tb20A1mu1phimu3}
illustrates similar studies with $A_f=1.5 ~{\rm TeV}$, $\mu=1$ ${\rm
  TeV}$ (left) and 
$A_f=1$ TeV, $\mu=1.5$ TeV (right), in particular 
showing how significantly
the result depends on $A_f$ and $\mu$. In general, we may 
expect the difference between the CP-conserving and CP-violating cases
to depend quite sensitively on $A_f$ and $\mu$, more so in the
presence of a light stop. 

Therefore, we conclude that 
the discovery of a light MSSM Higgs boson
(with mass below 130 GeV or so) at the LHC may eventually
enable one to disentangle the CP-violating case from the CP-conserving one, 
so long that the relevant SUSY parameters entering 
$gg\to H_1 \to\gamma\gamma$ are measured elsewhere,
in particular $m_{\tilde t_1}$. This is not
phenomenologically unconceivable, as 
this Higgs detection mode requires a very high
luminosity, unlike the discovery of those sparticles 
(and the measurement of their masses
and couplings) that enter the loops. 

\section*{Acknowledgments}
S. Munir's research is sponsored by DGAPA, UNAM, and CONACyT, Mexico, project no. 82291-F. SM is financially supported in part by the scheme `Visiting Professor
- Azione D - Atto Integrativo tra la Regione Piemonte e gli Atenei Piemontesi.

\end{document}